\documentclass{camera}
\usepackage{txfonts}   
\usepackage{relsize}
\usepackage{graphicx}
\usepackage{epsfig}
\RequirePackage{xspace}
\def\Vub  {\ensuremath{|V_{ub}|}\xspace}
\def\BB      {\ensuremath{B\Bbar}\xspace} 
\def\babar{\mbox{\slshape B\kern-0.1em{\smaller A}\kern-0.1em
    B\kern-0.1em{\smaller A\kern-0.2em R}}}
\def\cleo{CLEO}
\def\belle{BELLE}
\def\pep2{PEP-II}

\def\invfb   {\ensuremath{\mbox{\,fb}^{-1}}\xspace}
\mathchardef\Upsilon="7107
\def\Y#1S{\ensuremath{\Upsilon{(#1S)}}\xspace}
\def\FourS {\Y4S}

\def\Bb      {\ensuremath{\Bbar}\xspace}
\def\Bu      {\ensuremath{B^+}\xspace}
\def\Bub     {\ensuremath{B^-}\xspace}

\def\KL    {\ensuremath{K^0_{\scriptscriptstyle L}}\xspace} 
\def\Dbar    {\kern 0.2em\overline{\kern -0.2em D}{}\xspace}

\def\mes        {\mbox{$m_{\rm ES}$}\xspace}

\def\BzBzb   {\ensuremath{\Bz {\kern -0.16em \Bzb}}\xspace}
\def\BpBm    {\ensuremath{\Bu {\kern -0.16em \Bub}}\xspace}
\def\Bz      {\ensuremath{B^0}\xspace}
\def\Bzb     {\ensuremath{\Bbar^0}\xspace}

\def\BR         {{\ensuremath{\cal B}\xspace}}
\def\Bbar    {\kern 0.18em\overline{\kern -0.18em B}{}\xspace}

\newcommand {\rusl}{\ensuremath{R_{u/sl}}}

\newcommand {\Bxclnu}{\ensuremath{\Bb \rightarrow X_c \ell \bar{\nu}}}
\newcommand {\Bxulnu}{\ensuremath{\Bb \rightarrow X_u \ell \bar{\nu}}}

\newcommand {\mX}{\ensuremath{m_{X}}}
\newcommand {\Bxlnu}{\ensuremath{\Bb \rightarrow X \ell \bar{\nu}}}
\newcommand{\gevcc}{\ensuremath{{\mathrm{\,Ge\kern -0.1em V\!/}c^2}}\xspace}
\newcommand{\gevc}{\ensuremath{{\mathrm{\,Ge\kern -0.1em V\!/}c}}\xspace}
\newcommand{\mevcc}{\ensuremath{{\mathrm{\,Me\kern -0.1em V\!/}c^2}}\xspace}
\newcommand{\gev}{\ensuremath{\mathrm{\,Ge\kern -0.1em V}}\xspace}
\newcommand{\mev}{\ensuremath{\mathrm{\,Me\kern -0.1em V}}\xspace}

\newcommand{\mevc}{\ensuremath{{\mathrm{\,Me\kern -0.1em V\!/}c}}\xspace}

\def\btoc    {\ensuremath{b \to c}}
\def\bsg     {\ensuremath{b \to s \gamma}}
\def\btou    {\ensuremath{b \to u}}
\def\btoulnu    {\ensuremath{b \to u \ell \nu}}

\begin{document}

\title{$|V_{ub}|$ MEASUREMENTS AT $B-$FACTORIES }
\author{A Sarti}
%
\organization{Dep. of Physics, The University of Ferrara and INFN, I-44100 Ferrara, Italy}

\maketitle

\begin{abstract}
The determination of the \Vub~ element of the 
Cabibbo-Kobayashi-Maskawa (CKM) 
quark mixing matrix plays a central role in testing the Standard Model 
(SM) interpretation of CP violation.
Measurements at $B-$Factories are contributing with results 
that rely on different theoretical assumptions. Recent  inclusive and 
exclusive results from \cleo, \babar~ and \belle~ are reviewed. 
\end{abstract}

%


The element \Vub~ of the Cabibbo-Kobayashi-Maskawa (CKM)
matrix~\cite{ckm} plays a  central role in  tests  of the unitarity
of this matrix within the CP violation mechanism as predicted by
the Standard Model (SM). Precision \Vub~ measurements, based on
theoretical calculations performed at tree level, free of
new physics contribution expected in loop processes, can constrain
the SM predictions with boundaries that should survive any future SM
extension. \newline
Given that other main parameters of the CKM 
matrix are known with high precision, stringent SM tests require an 
uncertainty on \Vub ~ of less than 10\%. The present relative error, 
for both inclusive and exclusive measurement
techniques, is $\sim 15\%$. Measuring \Vub~ by using 
inclusive charmless semileptonic $B$ decays 
is a major experimental task due  to the very high \Bxclnu~ background
that needs to be rejected when  studying \Bxulnu~ decays (BR(\Bxclnu)
$\sim 60$ BR(\Bxulnu)). Kinematic variables describing the 
semileptonic decay, such as the lepton  energy ($E_{\ell}$), the
invariant mass of the lepton pair ($q^2$) 
or the invariant mass of hadronic system (\mX), can be used 
with different power to discriminate between \btoc~ and \btou~ 
transitions.\newline
Theoretically, the introduction of cuts for the background suppression
introduces additional errors due to the uncertainties on the 
parametrization of Fermi motion of the
$b-$quark inside the $B-$meson\cite{fazioneubert}.
The \Vub~ measurement using exclusive decays is also challenging from the
theoretical point of view: analyses deal with many signal models and Form
Factors (FF) lattice calculations.
One main experimental advantage of measurements done at $B$ factories is
the possible full reconstruction of a decaying $B$ meson that allows 
constraints on the missing momentum ($\nu$ reconstruction) and allows 
a $B$ meson clean  sample selection. Sample statistics is hence 
reduced, but cuts can be applied 
on charge conservation, $\nu$ reconstruction and lepton number, 
in order to clean up the sample from background events.   

\section{Inclusive analyses}

Two indipendent strategies are presented, based on the study of the $E_{\ell}$ 
endpoint spectrum and the \mX~ distribution. The first one 
is characterized by a large semileptonic events
reconstruction efficiency and low
signal events acceptance ($\sim$20\%). The latter has a lower
semileptonic events reconstruction efficiency, a large signal events
acceptance ($\sim$50-80\% of them are retained by a cut on \mX) and
a still competitive signal over background ratio (S/B).
Both approaches are mainly affected by the 
theoretical error on the extrapolation to full phase space.\newline
The study of the $E_{\ell}$ endpoint spectrum has been performed by 
\cleo \cite{CLEOend} and \babar \cite{BaBaend}. \cleo~ analysis is based
on a sample of 9.1\invfb~ on-peak (ON) and 4.3\invfb~ off-peak (OFF) data. 
Suppression of continuum background events has been performed using a 
neural network. The range of lepton momentum, that has been 
studied and computed in the Center of Mass frame, is ($p^*_{\ell}$) 
2.2-2.6(\gevc). \newline
This analysis uses the continuum background subtracted $p^*_{\ell}$ spectrum,
after the \BB~ background subtraction, 
to compute the differential branching ratio 
($\Delta B$), that can be studied in bins of $p^*_{\ell}$.
 \cleo~ partial branching ratio
result is $\Delta B$(2.2-2.6\gevc) = 0.230$\pm$0.015(stat)$\pm$0.035(sys).
The fraction of signal events ($f_{u}$) expected in a given bin of $p^*_{\ell}$
can be estimated by theoretical 
calculations making use of the \bsg~ photon energy spectrum. 
Using the differential BR and $f_u$ it is possible to compute the 
total \Bxulnu~  BR and extract \Vub~ according to 
\cite{pdg2002}, i.e. using the equation: 
\begin{equation}
\Vub =  0.00445 \sqrt{ (\frac{ B(\btoulnu) \cdot 1.55 ps } { 0.002 \cdot \tau_{B}
 } ) } \times (  1.0  \pm 0.020_{pert} \pm 0.052_{1/m_{b}^3} ).
\label{eq:vub_extr}
\end{equation}

\begin{figure}
\begin{centering}
\hbox{\hskip 0.5cm \epsfig{file=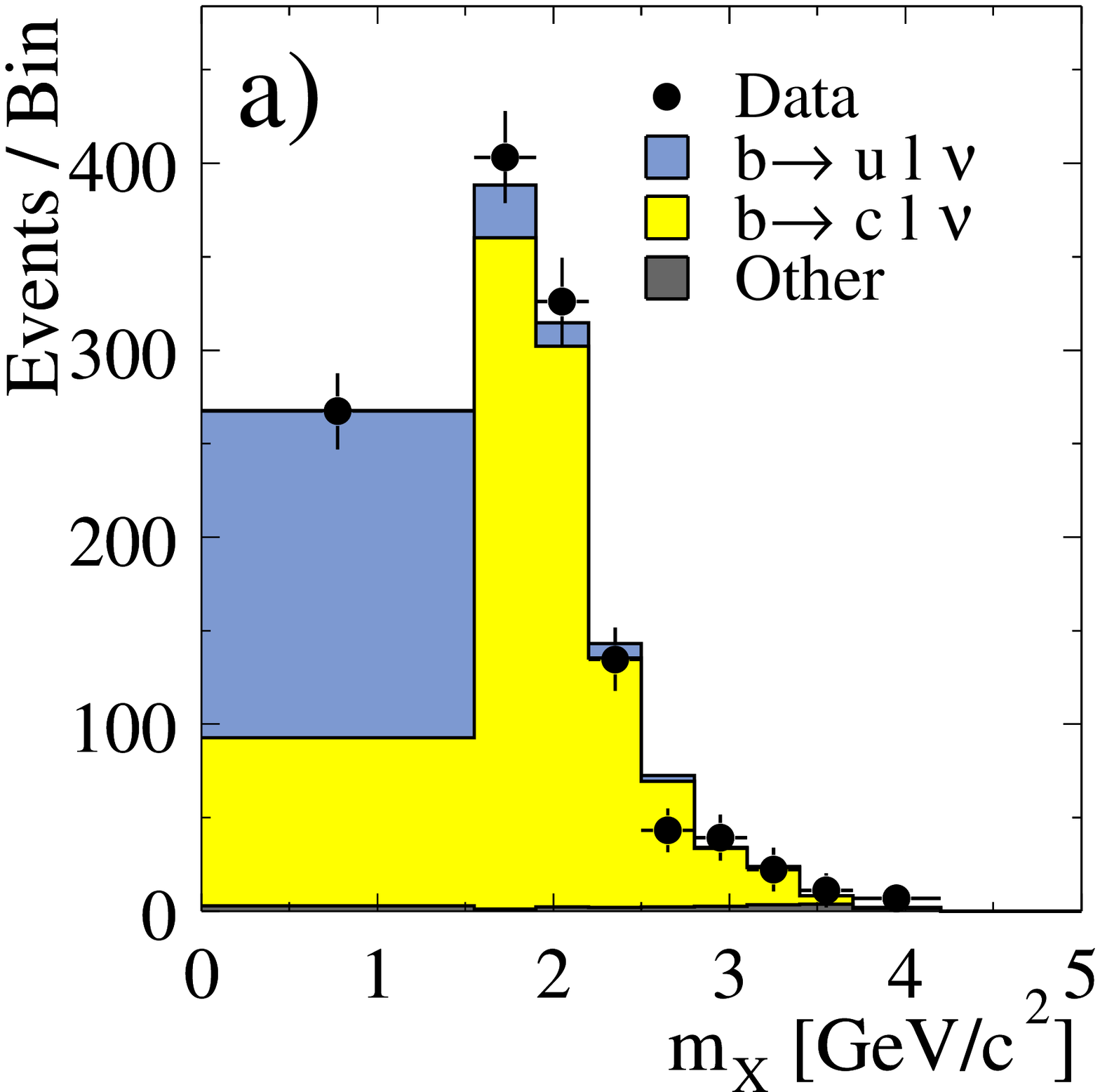,height=5.6cm} \epsfig{file=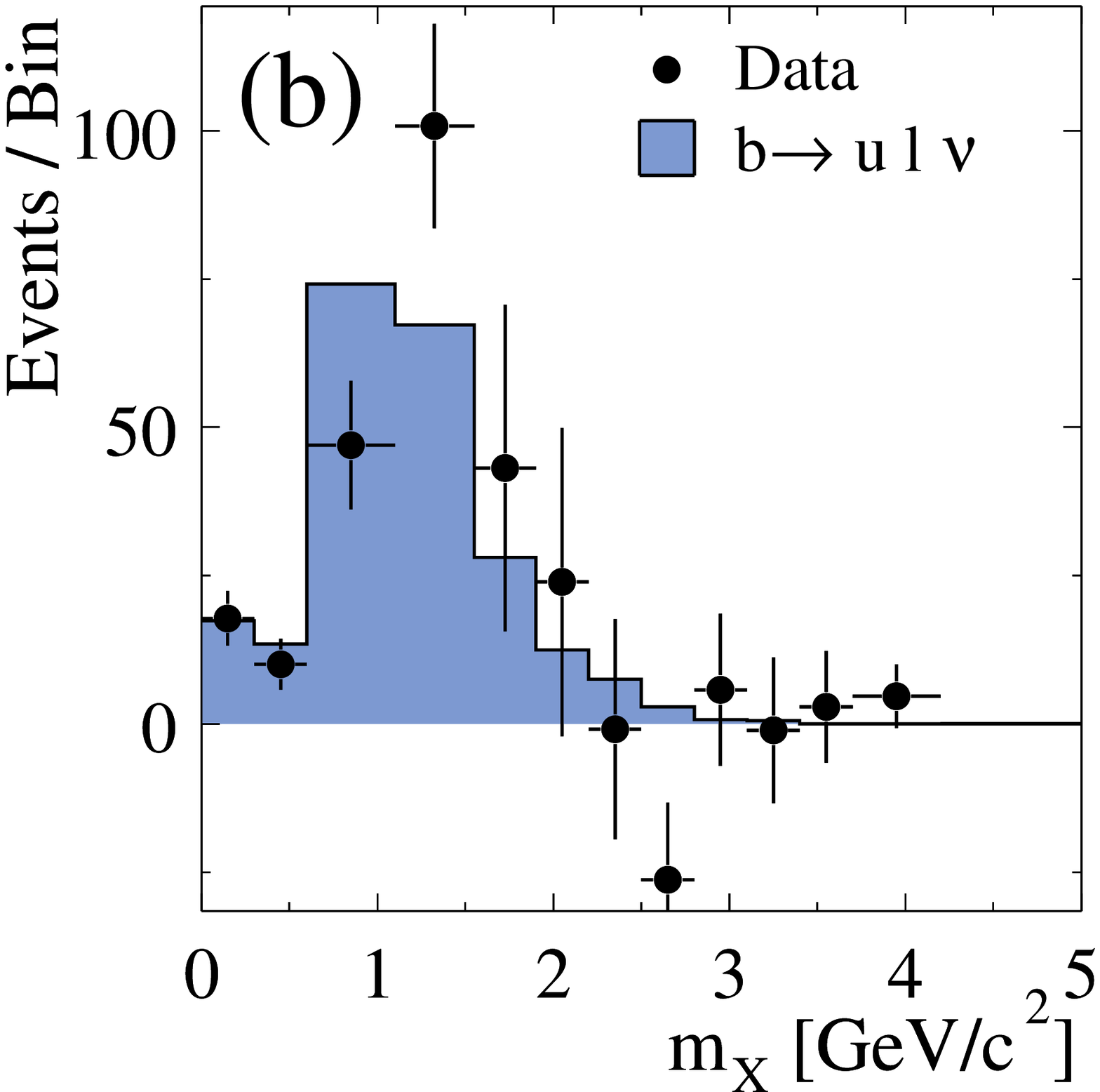,height=5.6cm} }
\caption{The $\mX$ distribution for \Bxlnu\ candidates for \babar \cite{babaincmx} 
inclusive analysis: a)
     data (points) and fit components, and b) 
     data and signal MC after subtraction of the $b\to c\ell\nu$ and the ``other'' backgrounds. }
\label{fig:mxspectra}
\end{centering}
\end{figure}

The result obtained by \cleo~ is $\Vub~ = (4.08 \pm 0.34 \pm 0.44 \pm 
0.16 \pm 0.24) \times 10^{-3}$. With a very similar approach \babar , in the
$p^*_{\ell}$ range 2.3-2.6(\gevc), obtains $\Vub~ = (4.43 \pm 0.29 \pm 
0.50 \pm 0.25 \pm 0.35) \times 10^{-3}$. Quoted errors are, 
respectively, the experimental one  (statistical plus detector 
systematic), the theoretical estimate of $f_u$,
the propagation of error in the above equation 
and the theoretical uncertainty on the validity of $f_u$
determination for \Bxulnu~ events using \bsg~ decays.\newline 
 The other inclusive method herein presented makes use of the
invariant mass of the hadronic system (\mX) in the \Bxlnu~ recoil
of a fully reconstructed B meson ($B_{reco}$) to separate the \btou~ 
and the \btoc~ transition. \babar~ analysis \cite{babaincmx}
reconstructs a large sample of  $B$ mesons by selecting hadronic
decays $B_{reco} \rightarrow  D^{(*)}X$. 
The kinematic consistency of $B_{reco}$ candidates 
is checked with two variables,
the beam energy-substituted mass $\mes = \sqrt{s/4 -
\vec{p}^{\,2}_B}$ and the energy difference 
$\Delta E = E_B - \sqrt{s}/2$. Here $\sqrt{s}$ is the total
energy in the \FourS center of mass frame, and $\vec{p}_B$ and $E_B$
denote the momentum and energy of the $B_{reco}$ candidate in the same
frame.  $\Delta E = 0$ is required within three standard
deviations as measured for each mode.\newline
On the recoil side, the event selection
proceeds via a lepton selection with a cut on the lepton momentum 
($\sim 1\gevc$) and on the missing mass of the event. Neutral 
and charged kaons are vetoed in the signal region and events with a kaon
positively identified in the recoil are
used as a control sample. \newline
In order to reduce systematic uncertainties,
the ratio of branching ratios $\rusl={\BR(\Bxulnu)/\BR(\Bxlnu)}$ 
is determined from $N_u$, the observed number of $\Bxulnu$ candidates 
with $\mX<1.55$\gevcc, and $N_{sl}$, the number of events with at 
least one charged lepton: 
\begin{equation}
\rusl=
\frac{\BR(\Bxulnu)}{\BR(\Bxlnu)}=
\frac{N_u/(\varepsilon_{sel}^u \varepsilon_{\mX}^u)}{N_{sl}} 
\times \frac{\varepsilon_l^{sl} \varepsilon_{reco}^{sl} } {\varepsilon_l^u \varepsilon_{reco}^u }.
\label{eq:vubExtr}
\end{equation}
Here  $\varepsilon^u_{sel}$ is the efficiency for selecting \Bxulnu\ decays
once a \Bxlnu\ candidate has been identified; $\varepsilon^u_{\mX}$ is the
fraction of signal events with $m_X < 1.55\gevcc$;
$\varepsilon_l^{sl}/\varepsilon_l^u$ corrects for the
difference in the efficiency of the lepton momentum cut for \Bxlnu\ and 
\Bxulnu\ decays, and $\varepsilon_{reco}^{sl}/\varepsilon_{reco}^u$
accounts for a possible efficiency difference in the $B_{reco}$
reconstruction in events with \Bxlnu\ and \Bxulnu\ decays.
$N_{sl}$ is derived from a fit to the \mes distribution.
$N_u$ is extracted from the $\mX$ distribution by a minimum $\chi^2$
fit to the sum of three contributions: the signal, the background
$N_{c}$ from \Bxclnu, and a background of $<1\%$ from other sources 
(misidentified leptons, secondary $\tau$ and charm decays). 
Fig.~\ref{fig:mxspectra}a shows the fitted $\mX$ distribution.
Fig.~\ref{fig:mxspectra}b shows the $\mX$ distribution
after background subtraction. 
By using 82\invfb integrated luminosity on the $\Upsilon(4S)$ peak,
\babar~ obtains $\Vub~ = (4.62 \pm 0.28 \pm 0.27 \pm 0.40 \pm 0.26)
\times 10^{-3}$, where errors are statistical, detector systematic, 
theoretical model and propagation of error in eq.\ref{eq:vub_extr}. 
The S/B ratio is $~$1.7 (higher than any previous inclusive analysis) 
and the main error comes from the Fermi motion parametrization. 
In figure \ref{fig:diffrate}, left plot, inclusive results 
for \Vub~ measurement 
are summarized togheter with previous results from LEP
experiments.  

\section{Exclusive analyses}

\begin{figure}[t]
\centering
\leavevmode
\epsfxsize=2.3in
\epsfig{file=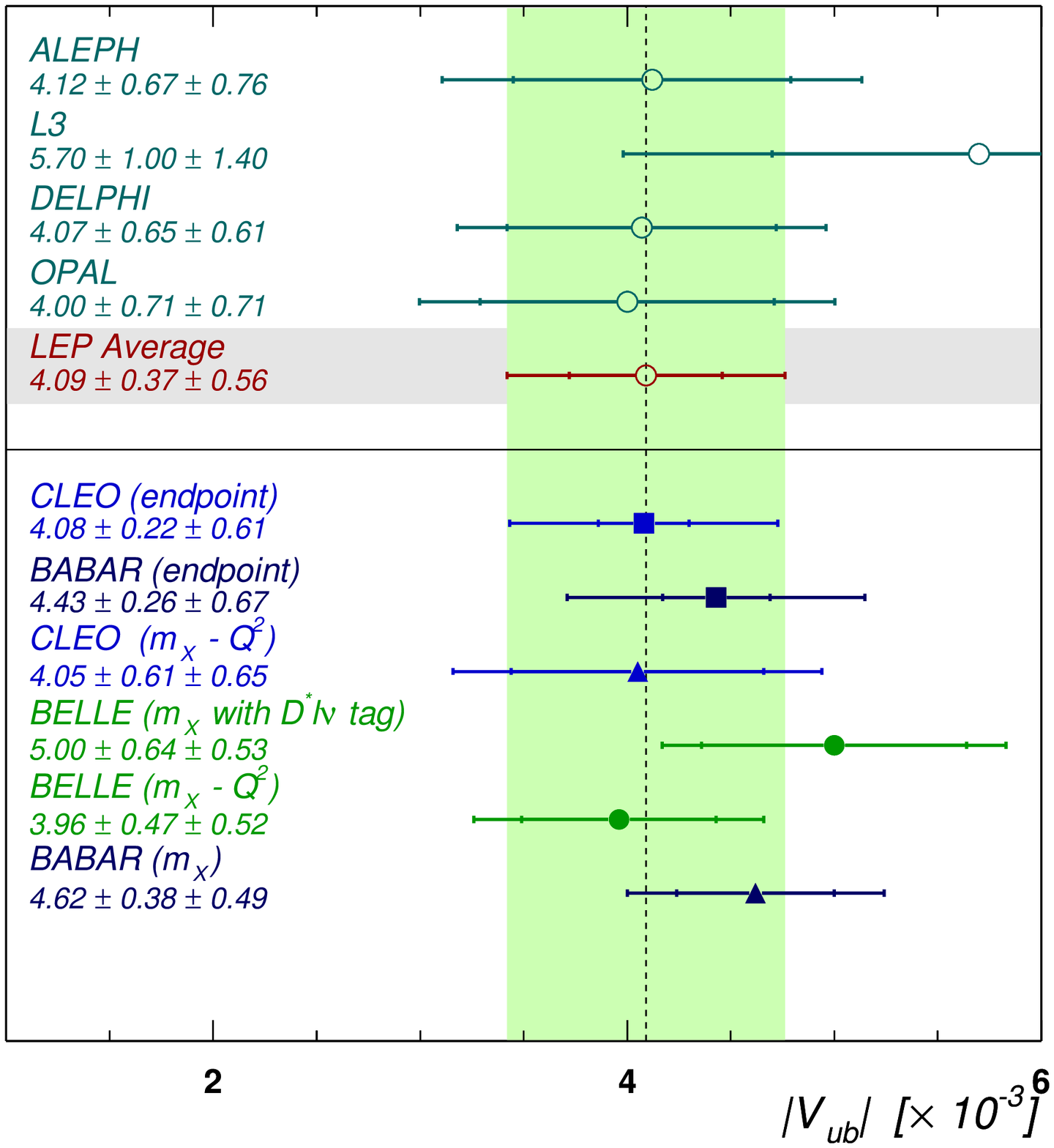,height=6.6cm} \epsfbox{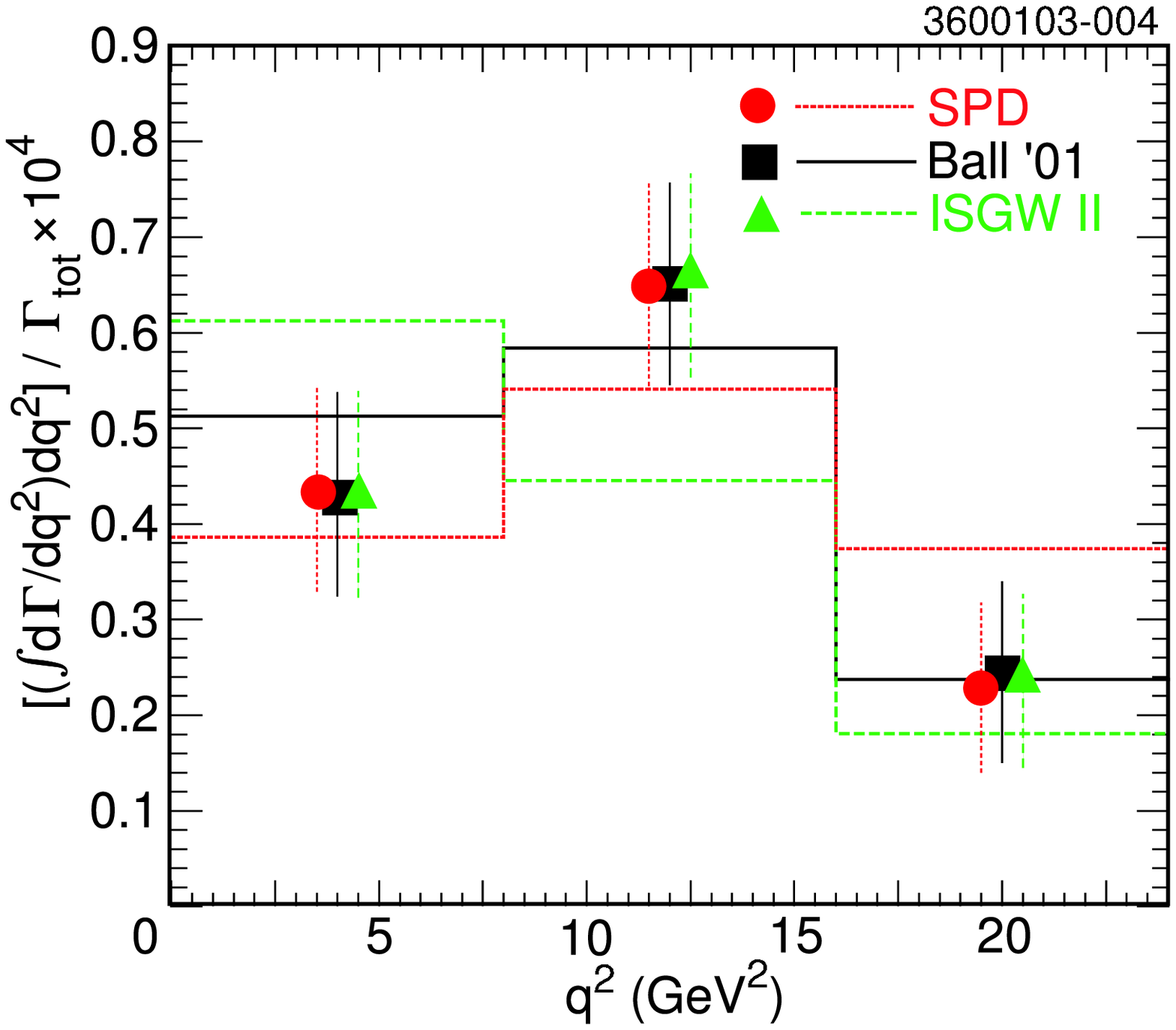}
\caption{Left: \Vub~ results from inclusive analyses. LEP and $B$ factories results are shown. Right: Measured branching fractions for $B^0\to\pi^-\ell^+\nu$
(points) and the best fit to the predicted $d\Gamma/dq^2$ (histograms) for the
three models used to extract both rates and $|V_{ub}|$.  The data points have 
small horizontal offsets introduced for clarity.}
\label{fig:diffrate}
\end{figure}

\babar \cite{babaex}, \cleo \cite{cleoex} and \belle \cite{belleex}
exclusive analysis strategies, based on the study of $B \rightarrow 
(\pi,\rho,\omega) \ell \nu$ decays, are outlined below. The signal events 
selection is done using the distributions of the invariant mass of 
$\pi\pi$ or $\pi\pi\pi$ system, the difference between the expected
and reconstructed $B$ meson energy ($\Delta E$) and lepton momentum 
($p_{\ell}$), for $B \rightarrow n\pi \ell \nu$ decays.
\cleo~ is using, for the $\rho$($\omega$) decay mode, a simultaneous 
fit to $m_{\pi\pi}(m_{\pi\pi\pi})$ and $\Delta E$ distributions.\newline 
The \cleo~ analysis achieves a low signal and detector modeling 
dependence by performing the full differential analysis in bins of $q^2$. 
The lepton momentum ranges covered are  p$_{\ell} >$ 1.0\gevc~ (pseudo-scalar
decay mode) and p$_{\ell} >$ 1.5\gevc (vector decay mode).
The main sources of systematics are: the modelization of \KL~ energy deposit,
the estimate of tracking efficiency, the simulation of charged and neutral 
particles interactions in the electromagnetic calorimeter and 
the physics model used to convert BR into \Vub .
On the simulation side,
signal modeling (various Form Factors for different exclusive decays)
and non-resonant contributions are playing a central role.
The efficiency for semileptonic and signal events selection, the 
continuum rejection and cross-feed background rates are depending on the event 
$q^2$, therefore possible bias can be
introduced by cuts on this variable and should be taken into
 account. A differential analysis that studies $d\Gamma/dq^2$ minimizes 
bias effects and allows to test different theoretical models.
Fig. \ref{fig:diffrate}, right plot, shows results 
for $B^0\to\pi^-\ell^+\nu$ decays: 
data (markers) and the best fit to the predicted $d\Gamma/dq^2$ (histograms)
for the three models used to extract both rates and $|V_{ub}|$ are shown. 
By improving precision it will be possible to
treat data with more appropriate theoretical models 
(for example in fig. \ref{fig:diffrate} the ISGWII model seems 
disfavored). 
\babar~ and \cleo~ have followed similar approaches analyzing different
p$_{\ell}$ ranges (p$_{\ell} >$ 2.3\gevc for \babar~ and $p_{\ell} >$ 2.4\gevc
for \belle). Table \ref{allexresu} shows the results for all 
the various exclusive analyses.
Attention must be paid when making averages: \cleo~ '03 analyses, for example,
are using different models for \Vub ~extraction with respect to the other
analyses and it is not clear how to take it properly
into account when averaging.

\begin{table}[ht] 
\begin{center}
\begin{tabular}{|c|c|}\hline
Analysis & Result \\ \hline
\cleo~  $(\rho)$    & $3.23\pm0.24^{+0.23}_{-0.26}\pm0.58$ \\
\babar~ $(\rho)$    & $3.64\pm0.22\pm0.25^{+0.39}_{-0.56}$ \\
\belle~ $(\pi)$     & $3.11\pm0.13\pm0.24\pm0.061$ \\
\belle~ $(\rho^0)$  & $3.50\pm0.20\pm0.28$ \\ \hline
\cleo~'03  $(\pi)$  & $3.24\pm0.22\pm0.13\pm0.09^{+0.55}_{-0.39}$\\
\cleo~'03  $(\rho)$ & $3.00\pm0.21^{+0.29}_{-0.35}\pm0.28^{+0.49}_{-0.38}$  \\
\cleo~'03  $(comb)$ & $3.17\pm0.17^{+0.16}_{-0.17}\pm0.03^{+0.53}_{-0.39}$  \\
\hline
\end{tabular}
\caption{Exclusive results for \Vub. Quoted errors are statistical, experimental systematic, theoretical, and signal Form Factor shape, respectively.}
\label{allexresu}
\end{center}
\end{table}

\section{Conclusions and outlook}

Several new results concerning the inclusive and exclusive determination 
of \Vub~ CKM matrix element have been obtained recently. The B-factory
era has introduced new experimental methods based on studies of the $q^2$ and 
mass spectrum in the recoil 
of fully reconstructed $B$ mesons.
Together with an impressive theoretical progress 
 (HQET and OPE for inclusive decays, FF determination for the 
exclusive ones) these improvements have lead to a 13\% precision
measurement (inclusive) and at a first attempt of discriminating
theoretical models via differential analysis (exclusive). The main open 
issues for inclusive analyses are the $\nu$ reconstruction and the
theoretical error reduction, while the exclusive ones are still suffering
from a poor FF determination. A \Vub~ measurement with an error less than
$10\%$ by 2007, and a consistency test for exclusive and inclusive 
determinations of \Vub~, the proper way to combine results being still under 
discussion, are the main goals regarding \Vub~ measurements in the near future.

%
\end{document}